\documentclass[twocolumn,showpacs,prl,amsmath,amssymb]{revtex4}
\usepackage{graphicx}
\usepackage{dcolumn}  
\usepackage{bm}       

\usepackage{latexsym}
\usepackage{amsmath}
\usepackage{amssymb}
\usepackage{enumerate}
\usepackage{bbm}
\usepackage[latin1]{inputenc}

\newcommand{\tr}{\textrm{Tr}}

\newcommand{\Eins}{\ensuremath{\mathbbm 1}}

\newcommand{\BE}{\begin{equation}}
\newcommand{\EE}{\end{equation}}
\newcommand{\kommentar}[1]{}

\newcommand{\mean}[1]{\ensuremath{\langle #1 \rangle}}
\newcommand{\qed}{\ensuremath{\hfill \Box}}
\newcommand{\be}{\begin{equation}}
\newcommand{\ee}{\end{equation}}

\begin{document}

\title{ Entanglement criterion based on skew information  }
\author{Zhihao Ma}
 \affiliation{Department of Mathematics, Shanghai Jiaotong
University, Shanghai, 200240, P.R.China, Email:
mazhihao@sjtu.edu.cn}

\date{\today}
\begin{abstract}
we can establish an entanglement criteria from skew information. our
criterion is independent of Local Uncertainty Relations (LUR)
([3],[4]).
\end{abstract}
\maketitle

The following can be found in [1],[3],[4]:

We start with the following observation. Let $\varrho$ be a density
matrix, and let $M$ be an observable. We denote the expectation
value of $M$ by $\mean{M}_{\varrho} := Tr(\varrho M)$ and the
variance (or uncertainty) of $M$ by
\begin{equation}
\delta^2(M)_{\varrho} := \mean{(M-\mean{M}_{\varrho})^2}_{\varrho}
= \mean{M^2}_{\varrho}-\mean{M}^2_{\varrho}.
\end{equation}
We suppress the dependence on $\varrho$ in our notation, when
there is no risk of confusion. If $\varrho$ is a pure state
the variance is zero iff $\varrho$ is an eigenstate
of $M.$ Now we have:

{\bf Lemma 1 of [1].} Let $M_i$ be some observables and
$\varrho=\sum_k p_k \varrho_k$ be a convex combination ({\it i.e.}
$p_k\geq 0, \sum_k p_k =1$) of some states $\varrho_k$ within some
subset $S.$ Then
\begin{equation}
\sum_i \delta^2(M_i)_{\varrho}
\geq
\sum_k p_k \sum_i \delta^2(M_i)_{\varrho_k}
\label{l1a}
\end{equation}
holds.

In this paper, we will discus new Relations, and get new
entanglement criteria.

First, note that Wigner and Yanase  introduced the following concept
in [2], skew information, it was defined as:
\begin{equation}
I(\varrho, M) :=\tr(\varrho M^{2})-\tr(\varrho^{\frac{1}{2}}M
\varrho^{\frac{1}{2}}M).
\end{equation}

where $M$ is an observable.

When $\varrho$ is a pure state, then $I(\varrho, M)$ reduce to the
variance $\delta^2(M)_{\varrho}$.

$I(\varrho, M)$ has the following celebrated properties:

(a): convex, i.e.,
\begin{equation}I(\lambda_{1}\varrho_{1}+\lambda_{2}\varrho_{2}, M)\leq
\lambda_{1}I(\varrho_{1}, M)+\lambda_{2}I(\varrho_{2},
M)\end{equation} for $\lambda_{1}+\lambda_{2}=1,\lambda_{1}\geq 0,
\lambda_{2}\geq 0$.

(b):
\begin{equation}I(\varrho_{1}\otimes\varrho_{2},
M_{1}\otimes I+I\otimes M_{2})= I(\varrho_{1}, M_{1})+I(\varrho_{2},
M_{2})\end{equation} where $M_{1}$ is an observable of Alice,
$M_{2}$ is an observable of Bob.

We have the following inequality:

{\bf Lemma 2.} Let $M_i$ be some observables and $\varrho=\sum_k p_k
\varrho_k$ be a convex combination ({\it i.e.} $p_k\geq 0, \sum_k
p_k =1$) of some states $\varrho_k$ within some subset $S.$ Then
\begin{equation}
\sum_i I(\varrho, M_i)\leq \sum_k p_k \sum_i I(\varrho_k, M_i)
\label{l1a}
\end{equation}
holds. We call a state ``violating Lemma 2'' iff there are no states
$\varrho_k \in S$ and no $p_k$ such that Eq. (\ref{l1a}) is
fulfilled.

\emph{Proof.} From property (a), we know that the inequality holds
for each $M_i$: $I(\varrho, M_i) = I(\sum_k p_k \varrho_k, M_i) \leq
\sum_k p_k I(\varrho_k, M_i).$ $\qed$

We know that $I(\varrho, M)$ and uncertainty $\delta^2(M)_{\varrho}$
coincides on pure state, i.e, when $\varrho$ is a pure state, then
the inequalities (2) and (6) in lemma 1, 2  both become equality.

let us recall the ``Local Uncertainty Relations'' (LURs), introduced
by Hofmann and Takeuchi [4]. Let $A_i$ be observables on Alice's
space of a bipartite system. If they do not share a common
eigenstate, there is a number $C_A > 0$ such that $\sum_i
\delta^2(A_i)_{\varrho_A}\geq C_A$ holds for all states $\varrho_A$
on Alice's space. Hofmann and Takeuchi showed:

{\bf Proposition 1. [4]} Let $\varrho$ be separable and let
$A_i,B_i, i=1,...,n$  be operators on Alice's (resp. Bob's) space,
fulfilling $\sum_{i=1}^n \delta^2(A_i)_{\varrho_A}\geq C_A$ and
$\sum_{i=1}^n \delta^2(B_i)_{\varrho_B}\geq C_B.$ We define $M_i :=
A_i \otimes \Eins +\Eins \otimes B_i.$ Then
\begin{equation}
\sum_{i=1}^n\delta^2(M_i)_{\varrho} \geq C_A + C_B
\end{equation}
holds.

The LURs provide strong criteria  which can by construction be
implemented with local measurements.

We have a dual result of the above Proposition of [3]:

{\bf Theorem 1. } Let $\varrho$ be separable and let $A_i,B_i,
i=1,...,n$ be operators on Alice's (resp. Bob's) space, fulfilling
$\sum_{i=1}^n I(\varrho_A, A_i)\leq C_A$ and $\sum_{i=1}^n
I(\varrho_B, B_i)\leq C_B.$ We define $M_i := A_i \otimes \Eins
+\Eins \otimes B_i.$ Then
\begin{equation}
\sum_{i=1}^n I(\varrho, M_i)\leq C_A + C_B
\end{equation}
holds.

\emph{Proof.} From property (b), we know that for product states,
$I(\varrho_A \otimes \varrho_B, M_i)=I(\varrho_A, A_i)+I(\varrho_B,
B_i)$, and from property (a), we know that after mixture,  the
inequality (8) holds $\qed$

The CCN criterion can be formulated in the following: see [3]. It
makes use of the Schmidt decomposition in operator space. Due to
that, any density matrix $\rho$ can be written as
\begin{eqnarray}
\rho=\sum_k \lambda_k G^A_k \otimes G^B_k. \label{rhodecompose}
\end{eqnarray}
where the $\lambda_k \geq 0$ and $G^A_k$ and $G^B_k$ are orthogonal
bases of the observable spaces,  Such a basis consists of $d^2$
observables which have to fulfill
\begin{eqnarray}
Tr(G^A_k G^A_l) = Tr(G^B_k G^B_l)=\delta_{kl}.
\label{localorthogonal}
\end{eqnarray}
We refer to such observables as {\it local orthogonal observables}
(LOOs). For instance, for qubits the (appropriately normalized)
Pauli matrices together with the identity form a set of LOOs.

{\bf Theorem 2.} To connect our criterion with  the LURs, first note
that for any LOOs $G^A_k$ the following relation \begin{equation}
\sum_{k=1}^{d^2} I (G^A_k) \leq d-1\end{equation} holds for any
states for Alice(Bob).

{\bf Proof.} Since  $\sum_k \tr(\rho (G^A_k)^2) = d \openone$ and
that $\sum_k \tr(\rho^{1/2}G^A_k \rho^{1/2}G^A_k)=
\tr(\rho^{1/2})^{2}\geq 1.$ $\qed$

 In [3], the authors get the following:
for separable states \be 1-\sum_k \mean {G^A_k \otimes G^B_k} -
\frac{1}{2} \sum_k \mean {G^A_k \otimes \openone  - \openone \otimes
G^B_k}^2 \geq 0. \label{lur3} \ee

For our criterion,  since $I(\varrho, M) :=\tr(\varrho
M^{2})-\tr(\varrho^{\frac{1}{2}}M \varrho^{\frac{1}{2}}M)$, so
Combining Eq (11) with the method of [3], using the fact that
$\sum_k (G^A_k)^2 = \sum_k (G^B_k)^2 = d \openone$  we can repair
the above inequality as: for separable states \be 1-\sum_k \mean
{G^A_k \otimes G^B_k} - \frac{1}{2} \sum_k
\tr(\varrho^{\frac{1}{2}}M_{k} \varrho^{\frac{1}{2}}M_{k}) \leq 0.
\label{lur3} \ee where $M_{k}:=G^A_k \otimes \openone  - \openone
\otimes G^B_k$.

In [3], it was proved that  Any state which violates the computable
cross norm criterion can be detected by a local uncertainty
relation, while the converse is not true.

Numberical experiment  show that the our criterion is independent of
the LURs.

\vskip 0.3 in

{\bf REFERENCES }

\vskip 0.2 in

[1]. Phys. Rev. Lett. 92, 117903 (2004)

[2]. Proc. Natl. Acad. Sci. U.S.A. 49, 910 (1963).

[3].  Phys. Rev. A 74, 010301(R) (2006)

[4]. Phys. Rev. A {\bf 68}, 032103 (2003).

\end{document}